# CPuS-IoT : A Cyber-Physical Microservice and IoT-based Framework for Manufacturing Assembly Systems

Kleanthis Thramboulidis, Danai C. Vachtsevanou, Ioanna Kontou
*Electrical and Computer Engineering, University of Patras, 26500 Greece*

**Abstract:** Today's customers are characterized by individual requirements that lead the manufacturing industry to increased product variety and volume reduction. Manufacturing systems and more specifically assembly systems (ASs) should allow quick adaptation of manufacturing assets so as to respond to the evolving market requirements that lead to mass customization. Meanwhile, the manufacturing era is changing due to the fourth industrial revolution, i.e., Industry 4.0, that will change the traditional manufacturing environment to an IoT-based one. In this context, this paper introduces the concept of cyber-physical microservice in the Manufacturing and the ASs domain and presents the Cyber-Physical microservice and IoT-based (CPuS-IoT) framework. The CPuS-IoT framework exploits the benefits of the microservice architectural style and the IoT technologies, but also utilizes the existing in this domain huge investment based on traditional technologies, to support the life cycle of evolvable ASs in the age of Industry 4.0. It provides a solid basis to capture domain knowledge that is used by a model-driven engineering (MDE) approach to semi-automate the development, evolution and operation of ASs, as well as, to establish a common vocabulary for assembly system experts and IoT ones. The CPuS-IoT approach and framework effectively combines MDE with IoT and the microservice architectural paradigm. A case study for the assembly of an everyday life product is adopted to demonstrate the approach even to non-experts of this domain.

## 1. INTRODUCTION

The 4th Industrial revolution has a tremendous impact on the society and the Internet of Things (IoT) plays a key role in this evolution (Bi *et al.* 2014). IoT, along with big data and cloud computing will allow the industry to cope with system complexity, increase information visibility and improve production performance (Yang *et al.* 2019). Manufacturing systems including Assembly Systems (ASs) are greatly influenced by these technologies and it is expected that very soon the IoT-based manufacturing environment will be a reality. However, the investment in traditional technologies, as for example IEC61131 based systems, is huge and there is a need for systems and components that have been developed based on the conventional approach to be integrated and exploited in the new IoT-based environment. Moreover, the adoption of IoT technologies in the manufacturing domain will greatly affect the development and operation processes of systems in this domain. Industrial engineers are not familiar with IoT technologies, which, when adopted, make the development process too complicated for them. Furthermore, there are additional challenges that industry faces (Erol *et al.* 2016), such as the need to switch from mass production to mass customization and the strong demand for real-time response at the machine control level.

The importance of digital assembly as a key component in manufacturing for assembly systems has been identified by several researchers, e.g., Xu *et al.* (2012). As Battaïa *et al.* (2018) claim, new technologies not only open new opportunities for the assembly systems, but also bring additional challenges to unleash these opportunities. Therefore, industry and academia are looking for new architectures, methodologies and tools to address the challenges in this domain, (Riedl *et al. 2014)*. One such architecture is the service-oriented architecture (SOA), which has attracted the interest of research and practitioners from the manufacturing domain since a long time ago (Cucinotta *et al.*, 2009). However, in practice the adoption of research results on SOA is not the expected one. The manufacturing industry is conservative and is expecting for a technology to reach an acceptable level of maturity before its adoption. During that time, a new paradigm based on the concept of microservice appeared and promises to change the way in which software is perceived, conceived and designed (Dragoni *et al.,* 2017). Microservices are the building block of the microservice architecture, that is one of the latest architectural trends in software engineering, promising to address several open issues in software development (Thönes 2015). The microservice architectural style is becoming popular and has recently been adopted by various large companies; it has already attracted the interest of the research community in the domain of manufacturing systems (Thramboulidis et al. 2018a). It has been acknowledged (Fortino et al. 2017) that Services will represent one of the real drivers for industrial IoT. For example, Casadei et al. (2019) introduce the term Opportunistic cyberphysical service to refer to cyber-physical services.

### 1.1. The CPuS-IoT approach

In this context, Thramboulidis *et al.* (2018a) have adopted the model-driven engineering (MDE) approach and extended it to exploit IoT technologies for the automation of various tasks of the development and operation phases of the AS. The framework they present exploits the benefits of the IoT technologies but also utilizes the existing huge investment in manufacturing that is based on traditional technologies. Authors claim that the microservice paradigm will have a significant impact on the way future manufacturing systems will be developed. They propose the integration of IoT technologies with the microservice architecture and examine alternative scenarios for their exploitation in manufacturing systems. The framework they describe exploits both technologies, i.e., microservices and IoT, and has the Cyber-Physical microservice (CPuS) as the key construct for the modelling of the cyber-physical manufacturing systems.

In this work, which is an extension of Thramboulidis *et al.* (2018b), we adapt the above framework to the assembly systems domain and further expand it to capture domain knowledge of this domain. Thus, the presented in this work CPuS and IoT-based (CPuS-IoT) framework for ASs, considers the CPuS as the key construct for the modelling of the manufacturing assembly system. In addition, it utilizes IoT technologies as glue among its constituent components, as far as their software interfaces are concerned. Machine assembly workers as well as the other constituent parts of the assembly platform, such as workbench and assembly tools, are modelled as cyber-physical components that expose their properties as primitive CPuSs (p-CPuSs). p-CPuSs are described using web technologies and are available for discovery and use during the development time of the assembly system. They are available as well, during the system's operation, that leads to a flexible assembly system able to address the challenge of mass customization. Moreover, the modularity at the assembly process level, that is required to address mass customization needs, is increased by modelling the assembly processes using the microservice architecture. Composite CPuSs (c-CPuSs) are defined as compositions/mashups of p-CPuSs using either the orchestration or the choreography pattern and Client/Server and Publish/Subscribe IoT protocols.

Evolvability requirements are addressed by considering CPuSs as resources that can dynamically be reserved, used and released by the system without human intervention. The Resource Description Framework (RDF) (Brickley *et al.* 2014) is utilized to have a machine-readable specification for CPuS that the plant offers. RDF is also used to capture the domain knowledge in terms of models and meta-models which enrich the framework and allow the use of reasoners to support the assembly engineer in the design and operation of the system. MDE is used in this framework to address the complexity of the development process as well as to get the other benefits of this paradigm in the ASs domain. The presented framework also allows, through the adoption of the UML4IoT profile (Thramboulidis *et al.*, 2016), the integration of legacy components, since the investment in conventional technologies is huge.

### 1.2. Outline of the paper

Key concepts of the proposed framework are three meta-models on which the modelling of the evolvable AS is based. The first step of the proposed modelling approach is to construct the product's structural model (PSM) utilizing a meta-model that captures the key constructs for the structural modelling of the product, which is presented in Section 4.2. Based on this, the assembly process model (APrM) can be automatically generated utilizing the corresponding meta-model, i.e., the Assembly Process meta-model (APrMM), that is presented in Section 5.1. A two-step approach for the specification of the assembly process based on the concept of CPuS is described in Section 6. The initial assembly process model (Section 5.3) is independent of the configuration of the assembly plant. This model is then automatically refined to get the process's platform specific model (Section 6.1). A product from everyday life, the IKEA Gregor office chair (Section 2.2.) is used as a case study to demonstrate the effectiveness of the proposed approach.

The contribution of this paper is not to define assembly sequence and job assignment algorithms but to describe a CPuS and IoT-based approach and the corresponding framework for the next generation of ASs in the context of Industry 4.0. Assembly sequence and job assignment algorithms can be implemented using the infrastructure of the framework assuming their adaptation to a two-stage modelling process adopted in the framework. Even though this paper focuses on the Assembly domain the key concepts apply to Manufacturing systems in general.

The rest of this paper is structured as follows: Section 2 presents background information, the case study that is used as a running example in this paper and related work. The architecture of the system is presented in Section 3. Section 4 describes the key concepts for automating the generation of the assembly process. The modelling of the assembly process along with the two-step approach in its specification is discussed in Section 5. A prototype implementation of the CPuS-IoT framework is discussed in Section 6. Finally, the paper is concluded in the last section.

## 2. BACKGROUND AND RELATED WORK

### 2.1 The communication gap

There is an always increasing number of papers that deal with the exploitation of modern technologies such as cloud computing and IoT in the domain of manufacturing, but also in the sub-domain of ASs. However, there is a communication gap between experts of the manufacturing domain and IoT ones. This is evident by looking at the publications in each domain. Papers from the manufacturing domain deal with IoT just by using the term without any reference to specific technologies or concrete proposals on how to exploit IoT, as for example Wang *et al.* (2014), and Liu *et al.* (2017). Domain expertise without good understanding of the technology does not provide a feasible solution and vice versa. Wang *et al.* (2017) discuss the challenges involved in the generation of assembly plans and argue that the use of IoT and cloud computing helps to address complexity and reconfiguration requirements. However, apart from a general reference that IoT is used to interconnect modules of the AS, there is no concrete description of the use of IoT technologies to achieve this goal. Moreover, they describe a proposal based on a model template of product and flow charts to describe the assembly process. The proposed model template is presented as a class diagram in an unorthodox way, e.g., it captures the assembly concept in 7 different classes with different semantics. It represents the connection as a specialization of Assembly (the other specializations being Mating and Motion), while in another class diagram the connection appears to be a generalization of Mate and Joining.

On the other hand, papers from the IoT community refer to the manufacturing domain, but either without any specific proposal on how to exploit it or using toy examples, since the manufacturing domain knowledge is missing.

The CPuS-IoT framework presented in this work provides a solid basis to capture domain knowledge and establishes a common vocabulary for both communities, that will also facilitate the automation of the AS development. However, close collaboration between the two communities is highly required for the evolution of such a vocabulary.

### *2.2 The target system*

Our goal is to automate the development and evolution of the AS, as an artefact that: a) will accept assembly requests in the form of assembly jobs, as shown in Figure 1, and b) will have the knowledge to be self-transformed to an assembly system capable of performing the specific assembly job. The proposed CPuS-IoT framework can be considered as an attempt to address three of the main challenges in the domain of AS, as highlighted by Hu *et al.* (2011) which gives a review of state-of-the-art research in the areas of AS design, planning and operations in the presence of product variety. These challenges are a) the current assembly representations are considered limited in terms of the comprehensiveness of assembly information. Bill-of-Material (BOM) cannot directly represent the complex physical assembly processes and liaison graphs are not considered suitable for representing hierarchical functional structures, b) an assembly representation enabling interoperability across various locations and software platforms is required, and c) determination of all possible assembly sequences is required as this greatly affects the total design process of a product.

The IKEA Gregor Chair example assembly system is used as a case study to present the key concepts of the proposed framework and to demonstrate its applicability. Figure 2 presents the layout of a laboratory prototype assembly platform used to realize the assembly process of the Gregor office chair. Assembly operations are performed by three robots, i.e., R1, R2 and R3. Workbench 1 has three fixtures; each fixture Fi passes sequentially through the three positions, i.e., pos1, pos2 and pos3, for the assembly to be completed. R1 and R2 move on axis to work on pos1 and pos2 respectively. We do not claim that this is the optimal configuration of the assembly platform, but it can be used to demonstrate the proposed approach.

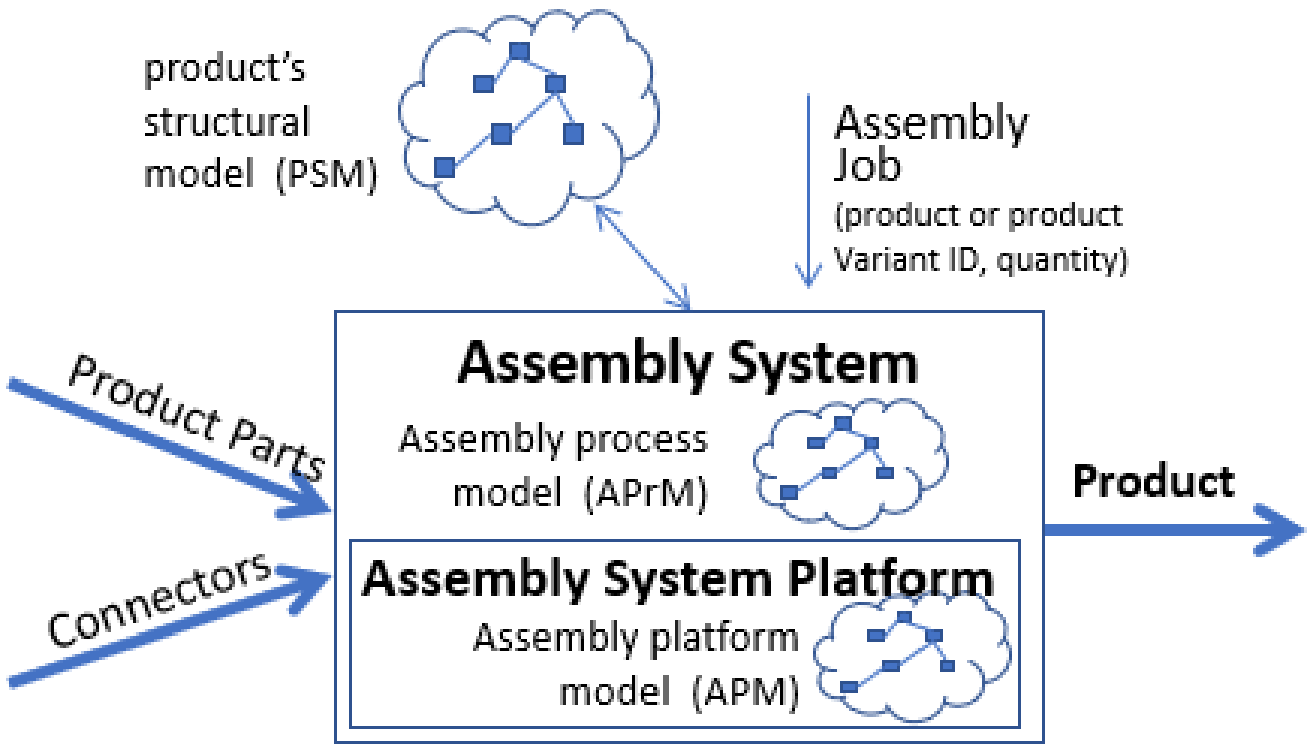

Fig. 1. The evolvable assembly system as considered by the CPuS-IoT framework.

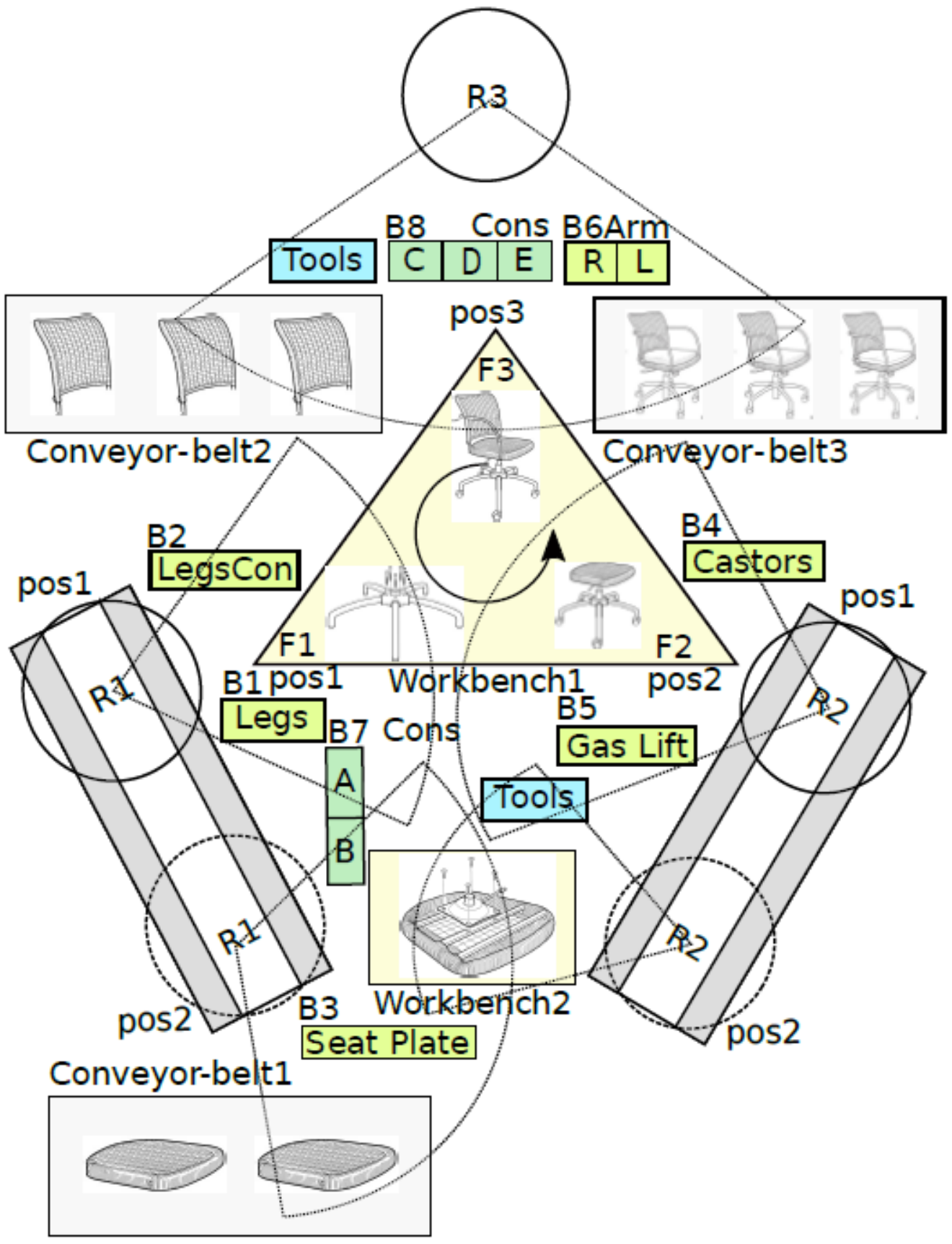

Fig. 2. The layout of the assembly platform as configured for the Gregor chair assembly process.

### *2.3 Assembly Systems and Industry 4.0*

Industry 4.0 represents the exploitation of enabling technologies such as the cyber-physical systems (CPS), IoT and cloud computing in the manufacturing industry (Xu *et al.* 2018). Many researchers are evaluating the evolution of Assembly systems in Industry 4.0, e.g., Cohen et al. 2017. A view regarding the design of ASs in the era of Industry 4.0 is presented by Bortolini et al. (2017a). Authors define balancing, sequencing, material feeding, equipment selection, learning effect and ergonomic risk as dimensions of AS design. Next they list the enabling technologies of Industry 4.0 that they consider as key players in the evolution of ASs.

IoT, real-time optimization, cloud computing, big data, machine learning and augmented reality are among them. They claim that the application of the IoT technology to assembly process is the keystone of the next generation of ASs which they call Assembly system 40 (AS40). Based on this infrastructure, they describe the main characteristics of an AS40 system. The entity, that they call assembly control system (ACS), leverages the available data to implement proper models and methods to automatically manage and configure the AS. In this context, our work can be considered as focusing on the exploitation of IoT technologies to propose a) an architecture for the IoT-based development of ASs, and, b) an approach and a framework to capture domain knowledge of the AS domain.

Manzini *et al.* (2018) present an approach for the design and reconfiguration of modular assembly systems through the integration of various computational tools in the automotive domain. The presented tool addresses the design of the system, the optimization of the layout, and the planning of reconfiguration actions and production. However, the presented approach does not exploit web technologies and IoT, not even provides a framework to capture assembly domain knowledge.

Thramboulidis (2016) presents an open distributed architecture for flexible hybrid ASs, based on the MDE approach. A model-based development process for development and operation of ASs is presented. IoT is described as a technology that will revolutionize the development and operation of ASs. The use of meta-models expressed in UML notation is proposed and it is claimed that the adoption of such an approach will drastically improve the development and operation of ASs. In this work, we go a step further; we extend the above work and utilize web technologies to represent the knowledge, captured in the various models of the framework presented by Thramboulidis (2016), in a form that can be processed by machines. The representation of the knowledge in a machine-readable way is a prerequisite for the exploitation of Web of Things (WoT) technologies to enable a further step in the automation of the design and operation of ASs. Meta-modelling has already been exploited to address the complexity of assembly systems. Xu *et al.* (2014), for example, describe a methodology based on object templates, which utilizes the so-called product assembly meta-model to address the complexity in product composition. In the CPuS approach, we extend the use of meta-models to address complexity in the assembly process and the assembly platform.

Ontologies have already been used by researchers in the AS domain, e.g., Sun *et al.* (2016) and Xu *et al.* (2014). Sun *et al.* (2016), for example, describe an ontology-based service model for CPSs and use an assembly line as a case study to demonstrate the usability of their model. They extend existing ontologies to satisfy the extra requirements in the modelling of services provided by CPSs. Our work is related to Sun *et al.* (2016) in several basic concepts regarding the description of the CPS services but it goes one step further by introducing the metamodels not only for the CPS service description but also for the product description as well as the assembly platform description.

Assembly sequence generation is a topic that has already been addressed by the research community, e.g., Sanderson *et al.* (1990) and Jones *et al.* (1997). Graphs is a tool that has been used in the generation process, e.g., Sanderson *et al.* (1990). We assume that the product designer is able to capture constraints and/or recommendations regarding the assembly process, independent of the approach adopted in the design of the product. i.e., concurrent assessment of assemblability during the product design phase or not. Based on this, we consider that the product designer, apart from the composition hierarchy information, has also significant information regarding the assembly process, as for example constraints on the assembly sequence. Thus, in the product's structural model, we capture information regarding the assembly process, such as suggested order to realize liaisons and master and branch sub-assemblies. Burnes *et al.* (2014) argue that the assemblability of a product is frequently neglected during the design phase of the product even though it is an important issue since it affects the partitioning of the product. They describe an approach that defines an assembly sequence concurrently with the product design. Moreover, several approaches have already proposed solutions for mass customization in the assembly domain, as for example Cecil *et al.* (2017) and Bortolini *et al.* (2017b).

To the best of our knowledge there is no other approach or framework that expoits the cyber-physical microservice along with IoT and MDE for the engineering of cyber-physical Manufacturing Systems.

## 3. ARCHITECTURE

In this section the architecture of the system as well as the architecture of the IoT Thing, which is used to represent the machine assembly worker in the system model, are presented.

### *3.1 System Architecture*

Manufacturing systems have been modelled for years based on the traditional five-layer (I/O, PLC, SCADA, MES, ERP) architecture which is also used in the ISA-95 standard. Layered architectures introduce several advantages but also introduce significant overhead in terms of performance. Modern technologies such as the Cloud and the IoT provide alternative solutions to this 5-layer architecture and are used by several organizations to promote a model where sensors send data directly to the cloud and services (e.g., production scheduling) automatically subscribe to necessary data in real-time, which is also, as claimed, the vision of cyber physical systems. Systems that adopt this model collect all raw events from sensors and process these on the Cloud. We do not adopt this view in our framework. Instead, we exploit edge and fog computing to process locally raw data produced by assembly workers and the other components of the assembly platform, as is also the case with Thramboulidis *et al. (2017).* This increases the overall responsiveness of the system and lowers its cost (*www.openfogconsortium.org*). This approach is the one adopted in the MIM model (Thramboulidis 2015) since, manufacturing systems require processing closer to the physical part, which avoids the introduction of unnecessary latency and decreases the likelihood of network failures (Thramboulidis *et al. 2017).*

Figure 3 captures the high-level architecture of the assembly system adopted in this work. The bottom layer of the architecture, i.e., the Primitive Cyber-Physical Microservice (p-CPuS) layer, consists of cyber-physical microservices. p-CPuSs encapsulate the mechanical units of the plant, i.e., the machine assemblers and the other mechanical units of the assembly platform, i.e., workbenches and tools. p-CPuSs transform these into IoT-enabled entities that provide assembly services to their environment. They encapsulate sensors, actuators and the low-level coordination logic required to offer more advanced functionality compared to the one offered by the mechanical unit. Optionally, the developer may export, at this level, properties of the mechanical unit exclusively for monitoring purposes. Services of the p-CPuS are used by c-CPuSs of the assembly process layer to implement the assembly processes. A similar term to the CPuS, i.e., the term cyber-physical service is used by Casadei et al. (2019). However, they define the cyber-physical service, which they call Opportunistic service, as an interface that allows an IoT Entity to be engaged in usage relationships.

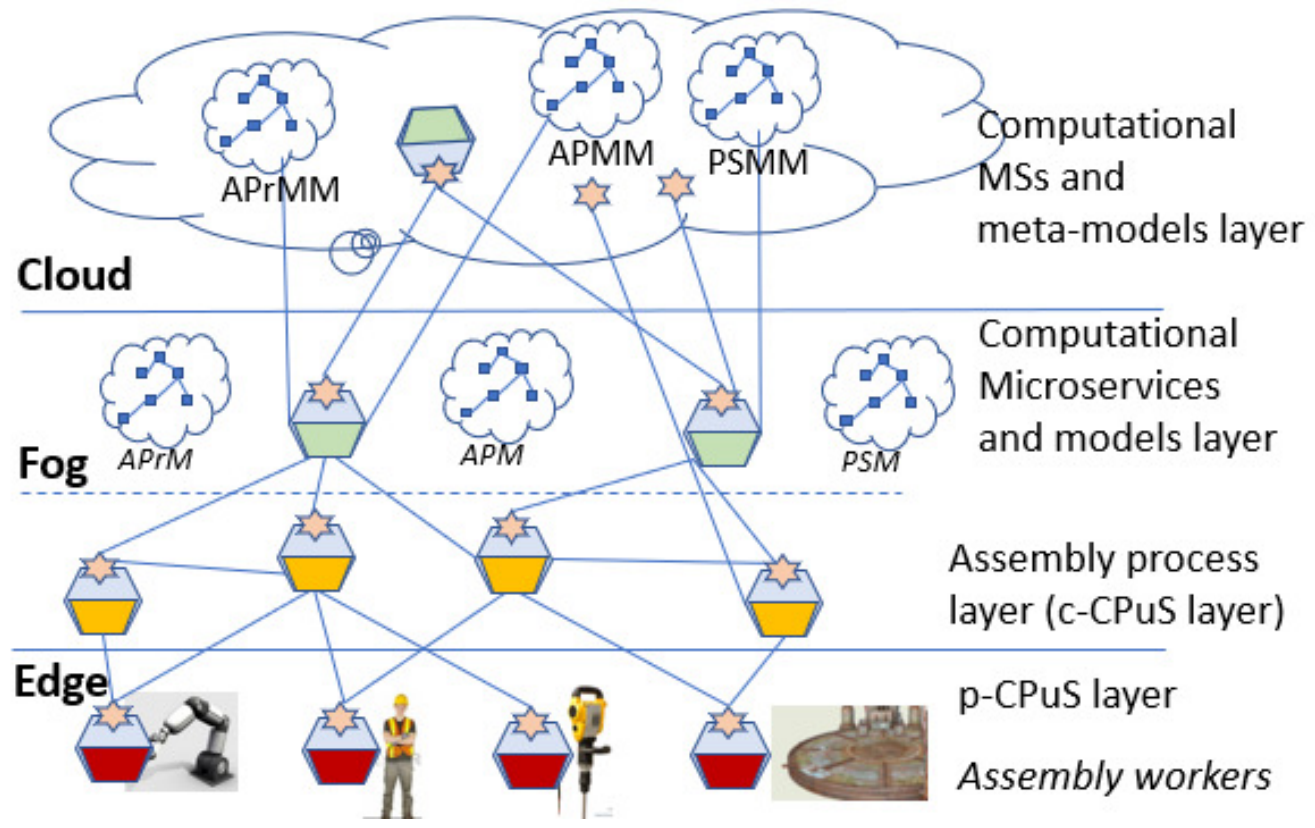

Fig. 3. The high-level architecture of the system.

The next layer, which is the fog layer, plays the role of private cloud. Assembly processes are deployed in the fog layer and are mainly defined as compositions/mashups of services provided by the assembly workers of the edge layer. Plant processes also utilize computational services offered by computational microservices, such as plant path generators, which are also deployed on the fog layer. Plant processes highly depend on the knowledge captured by the models of the framework. These models include, the Assembly Platform Model (APM), the Product's Structural Models (PSM) and the Assembly Process Model (APrM) as defined by Thramboulidis (2016), which are also deployed on the fog layer. It is suggested that the meta-models of the above models be deployed on the cloud, which is considered as the third layer of the architecture, so as to be widely available through the web to any AS. The PSM, APrM and APM are deployed on the fog for security reasons (Stergiou *et al.* 2018) since they capture intellectual property.

*3.2 Architecture of the IoT-compliant Assembly worker*

For the traditional machine assembly workers to be IoT-compliant their cyber interface should be transformed to a RESTful one. We have adopted the OMA LwM2M application layer protocol, which is implemented on top of CoAP, (an MQTT based implementation also exists) to provide an IoT-compliant interface for the machine assembly worker, as shown in Figure 4. IPSO objects were adopted to address interoperability requirements. We call IoT wrapper the software layer that transforms the legacy interface to an IoT-compliant. This wrapper transforms the conventional machine assembly worker to an IoT-compliant one, i.e., to an IoT Industrial Automation Thing. We found the adaptation process too complicated for the industrial engineer, which motivated us to use MDE to automate its construction.

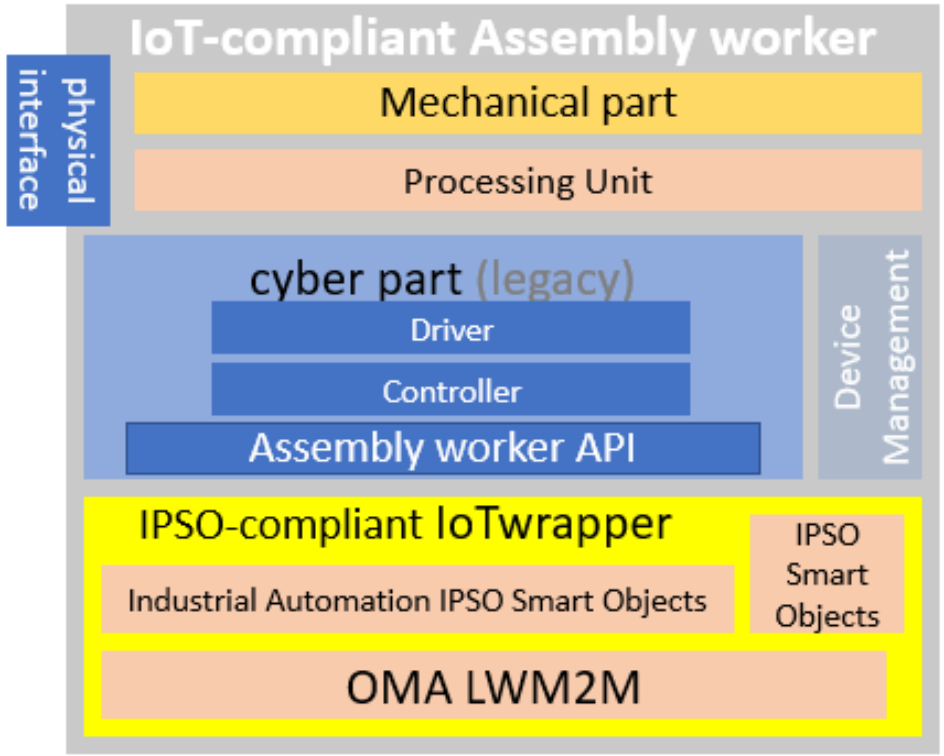

Fig. 4. Architecture of the IoT-compliant assembly worker.

For the specification of the IoT-compliant interface of the assembly worker, the LwM2M provides an object model that is based on the concept of Resource. This model focuses only on the modelling of the interface. On the other hand, the traditional assembly worker has been specified with an object model that also specifies its interfaces. UML and SysML, the de-facto standards for software and system engineering are commonly used for such a specification. Thus, we have two models; one focuses only on the IoT-compliant interface, and the other on the whole machine assembler including its interface, which cannot however be specified in an IoT-compliant way.

To address the above problem, we have defined the IoT-layer on top of the Cyber-Physical Microservice layer, that was defined to model the Assembly system as a composition of CPuSs, as shown in Figure 5. The modelling space of this layer is defined by a meta-model (Thramboulidis *et al.* 2017) which was constructed using the basic constructs of the LwM2M object model. In this way, projecting the CPuS layer model elements of the AS to the IoT-layer we get the IoT compliant interface for the constituent components of the AS, as well as, for the AS as a whole. UML was adopted as the base for the transformation process between the two layers, and a UML profile, the UML4IoT, was defined to implement this projection. The microservice architectural paradigm was adopted and adapted to the cyber-physical domain to provide flexibility for assembly workers and assembly processes.

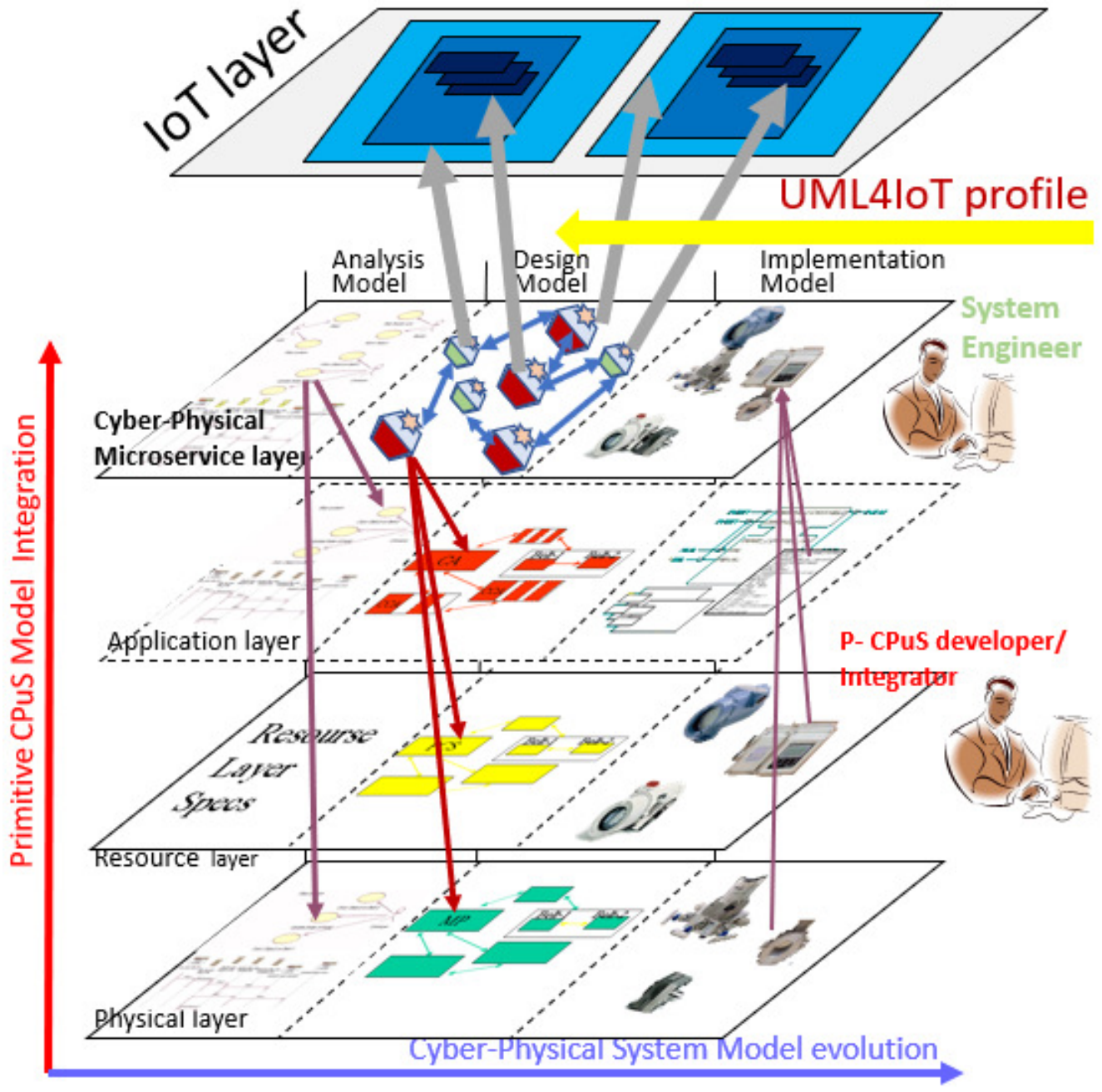

Fig. 5. Modelling the Assembly system as a composition of IoT-compliant CPuSs.

## 4. THE GENERATION OF THE ASSEMBLY PROCESS

During the past years many developments in ergonomics research and methods have been developed but Hierarchical Task Analysis (HTA) has remained a central approach (Santon 2006; Naweed *et al.* 2018). HTA-based approaches are widely used in assembly system design, e.g., (Mateus 2018; Tan *et al.* 2009). In this work, we follow a different approach for the modelling of the assembly process. The approach, which is characterized as bottom-up, is based on the product's structural model (PSM) and adopts two key models for the assembly process (APr). The first one is abstract and independent of the assembly platform and the second one is assembly platform specific. The key constructs for the modelling of both versions of the assembly process are captured by a meta-model, namely the Assembly Process Meta-Model (APrMM).

We do not consider BOM and liaison graphs as assembly representations, not even do we use these terms. Alternatively, we discriminate between structural and behavioural information and capture this knowledge in three separate meta-models, namely the Product's Structural Meta-Model (PSMM), the APrMM and the Assembly Platform Meta-Model (APMM). The APrMM along with the other two meta-models are used:

a) to formalize the domain knowledge and establish a common vocabulary between the AS community and the IoT one, and,

b) facilitate the development and operation of the AS.

Moreover, these meta-models act as a kind of domain-specific language to manage the complexity of the AS by effectively expressing domain specific concepts.

The objective of the CPuS-IoT framework is to automate the development and evolution of evolvable assembly systems. This is based on the modelling of the assembly platform, the assembly process and the target product. The models of the assembly platform and the product constitute the basis for the automatic construction of the assembly process defined by the CPuS-IoT approach.

### *4.1 The Assembly platform*

The assembly system platform is defined as a composition of IoT-compliant assembly workers that expose their properties as p-CPuSs represented as resources. The interface of the p-CPuS is modelled using UML provided and required interfaces. Provided interfaces are used to capture the assembly services provided by the assembly worker to its environment as CPuSs. CPuSs act as access points to trigger the execution of the corresponding assembly activities that the worker may execute. RDF is used to describe these resources and their relationships to represent the worker's model in a machine-readable format. For the platform description we consider that a service offered by an assembly worker manipulates physical objects in space and usually changes their state and location. For a physical object to be manipulated by the service, the object should be in a given location/region and a given state. For example, F2 of Workbench1 (see Figure 2) requests a service from R2 when the sub-assembly of Gregor chair in pos1 has been completed and is in pos2 after the rotation of W1. The term operation space is defined by Sun *et al.* (2016) to capture this information for the object manipulated by a service.

### *4.2 The product's structural meta-model*

Figure 6 presents the core of the product's structural meta-model expressed as UML profile. The product is considered as a composition of parts, which are either composite (*CompositePart*) or primitive (*PrimitivePart*), liaisons and optionally connectors. The Liaison is used to represent a mating relationship between parts in an assembly. Liaisons are classified as *SelfDefinedLiaison* or LowerDcl DefinedLiaison (*LoDclDefinedL*). The *SelfDefinedLiaison* is associated with 1 or more *LiaisonPair* and is used to represent a connection point of the specific mating that is defined by two *LiaisonEndPoint*. In this work, we abstract from our models the details that do not affect the approach described in this paper. Thus, we only model the connection points of the part, which we call liaison endpoints. The order of the liaison, which is imposed by the product's structure, and its type, which represents the specific connection among constituent parts of the product, are captured as properties of the liaison. These properties play a dominant role in the realization of the liaison. Liaison properties are classified and organized in several ways. For example, Barnes *et al.* (2004) define three attributes: Mating Joint Type, Assembly Action and Joining Process. For simplicity reasons, we only capture a few attributes in our model, that are required for the case study and the demonstration of the proposed approach. Further attributes may easily be captured in the meta-model.

For the demonstration of our approach, the defined by Swain *et al.* (2014) mapping of liaison types to assembly operations is adopted. Assembly operations are offered by corresponding services provided by the assembly workers.

This assumption is used to demonstrate our proposal for automatically constructing the APM from the product's structural model. However, the proposed approach is independent of the specific type of liaisons and the corresponding assembly operations. Other types of liaisons can also be used in the model, as for example the liaison types used by Loshe *et al.* (2005).

A product may have several sub-assemblies. It has a master sub-assembly (*MasterSubAssembly*) and optionally several branch sub-assemblies (*BranchSubAssembly*), with each one optionally having its own assembly line. A *SubAssembly* has one of its parts as its base part. This is captured by the association stereotype *HasAsBasePart*. In the Gregor chair case study, a *BranchSubAssembly* is defined and realized on W2. The *MasterSubAssembly* is realized on W1.

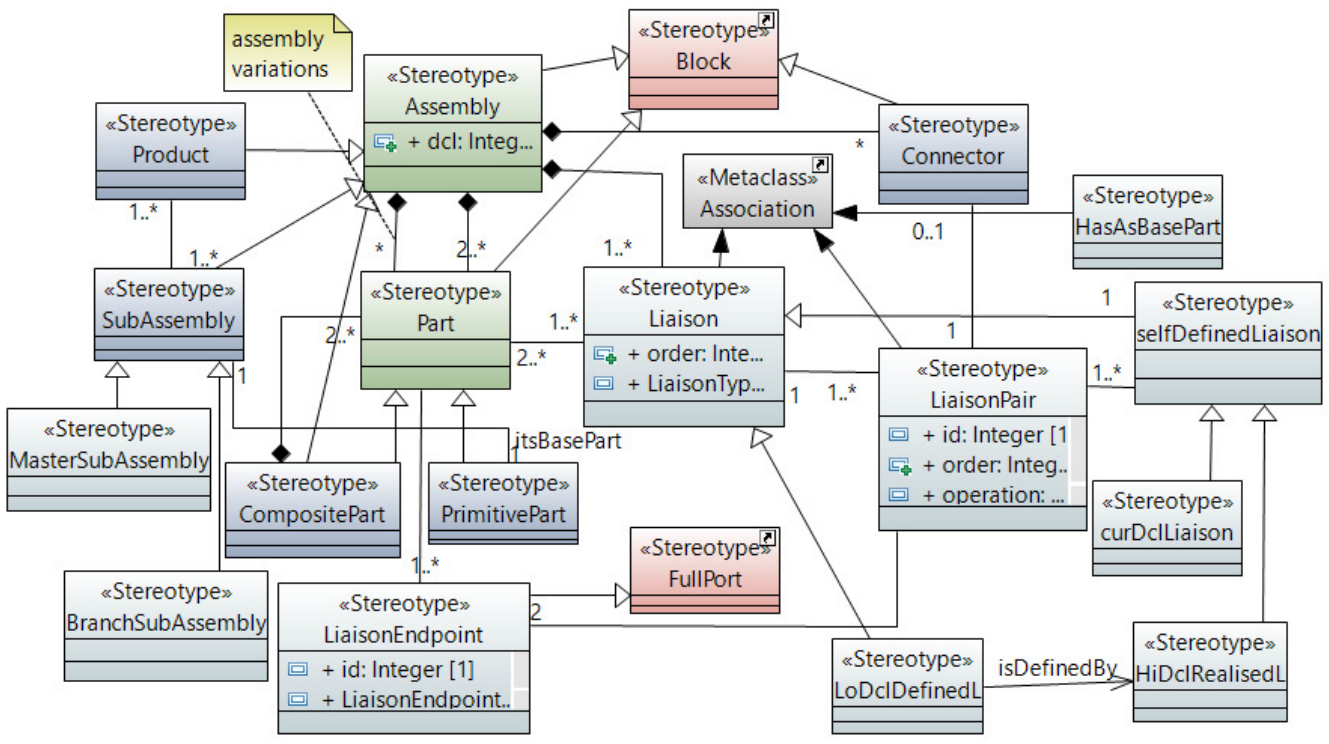

Fig. 6. The product's structural meta-model as a UML profile (core part).

The PSMM has been expressed as a UML profile that can be used by the AS Engineer to define the structural model of the product using a UML tool. The PSMM is also offered in a structured machine processable representation expressed in OWL DL (https://www.w3.org/TR/owl-guide/) and RDF In this case, the AS engineer should represent the product's structural model in RDF notation or generate it from the UML model. The RDF model can be used by inference engines for the construction of the APM. Figure 7 presents part of the RDF model of the PSMM.

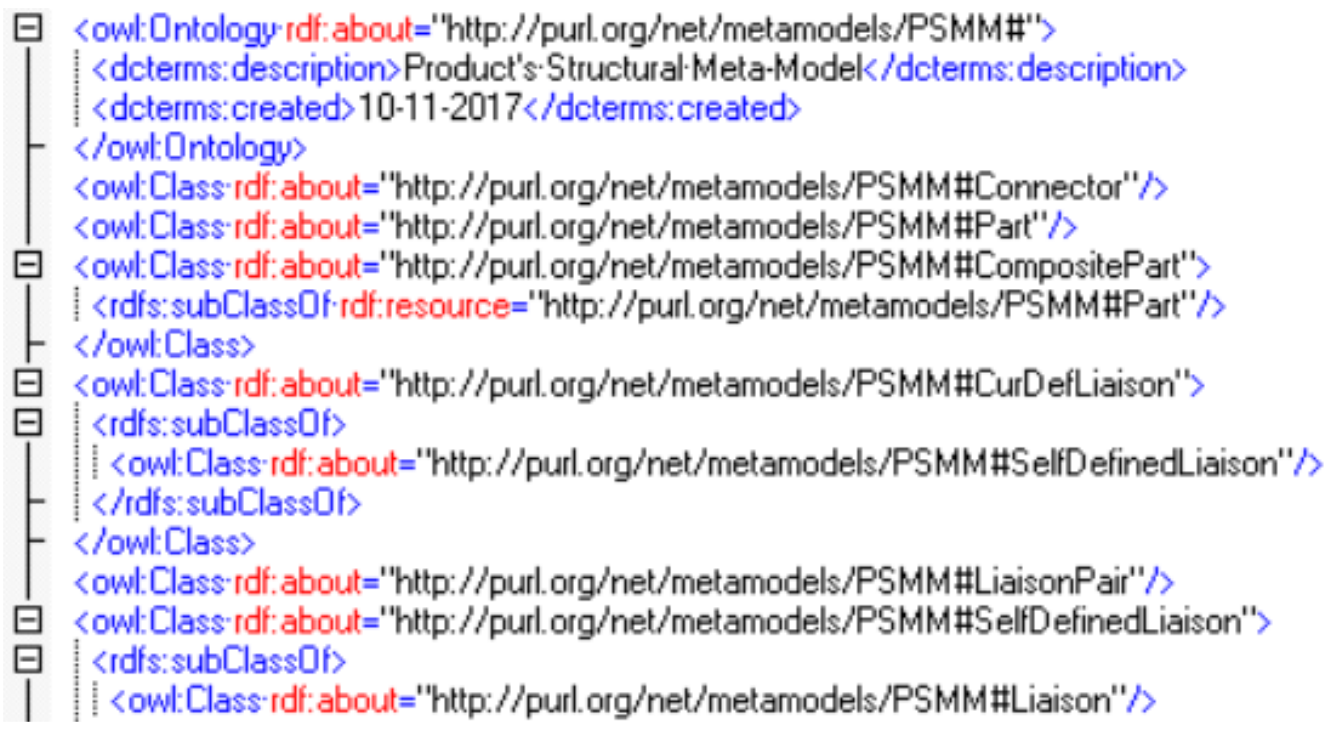

```
<owl:Ontology rdf:about="http://purl.org/net/metamodels/PSMM#">
  <dcterms:description>Product's Structural Meta-Model</dcterms:description>
  <dcterms:created>10-11-2017</dcterms:created>
</owl:Ontology>
<owl:Class rdf:about="http://purl.org/net/metamodels/PSMM#Connector"/>
<owl:Class rdf:about="http://purl.org/net/metamodels/PSMM#Part"/>
<owl:Class rdf:about="http://purl.org/net/metamodels/PSMM#CompositePart">
  <rdfs:subClassOf rdf:resource="http://purl.org/net/metamodels/PSMM#Part"/>
</owl:Class>
<owl:Class rdf:about="http://purl.org/net/metamodels/PSMM#CurDefLiaison">
  <rdfs:subClassOf>
    <owl:Class rdf:about="http://purl.org/net/metamodels/PSMM#SelfDefinedLiaison"/>
  </rdfs:subClassOf>
</owl:Class>
<owl:Class rdf:about="http://purl.org/net/metamodels/PSMM#LiaisonPair"/>
<owl:Class rdf:about="http://purl.org/net/metamodels/PSMM#SelfDefinedLiaison">
  <rdfs:subClassOf>
    <owl:Class rdf:about="http://purl.org/net/metamodels/PSMM#Liaison"/>
```

Fig. 7. RDF/XML description of the PSMM (part of).

## 5. TOWARDS A GOAL-DRIVEN ASSEMBLY PROCESS SPECIFICATION

### *5.1. The Assembly process meta-model*

The assembly process is specified, at the assembly platform level, as a composition of CPuSs provided by machine assemblers. Both service orchestration and service choreography patterns can be used for the definition of the assembly process. The assembly process of this level is highly dependent on the assembly platform. Thus, we call it platform-specific and its model assembly process platform-specific model (APr-PSM). To increase the reusability of the assembly domain knowledge and automate the generation process of the APr-PSM, another level of specification of the assembly process, that is independent of the assembly platform, has been defined. For the specification of this level of the assembly process, a model independent of the assembly platform, i.e., the APr-PIM, is used. To proceed with the definition of the APr-PIM, we considered the decomposition of the product into decomposition levels (dcl-i), and defined the assembly process as follows:

$$AP = \sum_{i_1=1}^{K} CCAP_{i_1} + \sum_{i_2=K+1}^{N} PCAP_{i_2} + \sum_{i_3=1}^{M} PAA_{i_3}$$

*where*
***N*** *is the number of parts at the first level of decomposition (dcl-0),* ***K*** *is the number of parts at dcl-0 that result to* ***CCAP*** *and* ***M*** *is the number of liaisons at dcl-0.*
*Composite-Child Assembly Process (**CCAP**) is the assembly process of a composite part which includes at least one composite part as constituent component,*
*Primitive-Child Assembly Process (**PCAP**) is the assembly process of a composite part whose all parts are primitive, and*
*Primitive Assembly Activity (**PAA**) is the assembly activity of realizing a liaison at the dcl-0 level of decomposition.*
*CCAP and PCAP are defined recursively using the same type.*

To assist the assembly engineer with the specification of the assembly process, a two-step approach is proposed. In the first step, the APr-PIM is generated. This model captures a) the chunks of functionalities that should be performed for the realization of the liaisons captured in the structural model of the product, and b) the precedence constraints among them that emanate from the structural model. In a next step, the APr-PIM is mapped to the assembly platform exploiting its model. The result is the APr-PSM which, along with the assembly platform, constitutes the AS for the specific product or its variants. Figure 8, which represents part of the APMM, captures the key modelling constructs for the specification of both versions of the assembly process, i.e., the APr-PIM and the APr-PSM. Both extend the *AssemblyProcess* which is defined to extend the SysML *Block* stereotype. The APr-PIM is composed of:

a) a set of Assembly Tasks (**AT**) which represents the work required to realize the liaisons of the PSM,
b) the Assembly Task Precedence Graph (**AT-PG),** that captures the precedence relations that exist among the ATs of the APr-PIM, and
c) the specifications, in machine-readable representation, of the assembly activities required for the realizations of the ATs.

As assembly task (*AssemblyTask*) we define the piece of the assembly work that is related to the realization of one or more liaisons of the APr-PIM. For the realization of an AT, a set of assembly activities (*AssemblyActivity*) should be performed with the main activities to result from the liaison types of its liaisons. These required assembly activities (*itsRequiredActivity)* are modelled as required services, i.e., as services that are required for the realization of the AT and should be mapped to services provided by assembly workers (provided services) during the transformation of the APr-PIM to APr-PSM. This transformation utilizes services (provided and required) as the primary decision criteria for the assignment of the assembly job to the assembly platform's workers, characterizing it as a service-oriented job assignment. A term commonly used in the assembly system domain is capability. For example, Ranz *et al.* (2017) describe a capability-oriented job assignment where capability is used as the primary decision criterion.

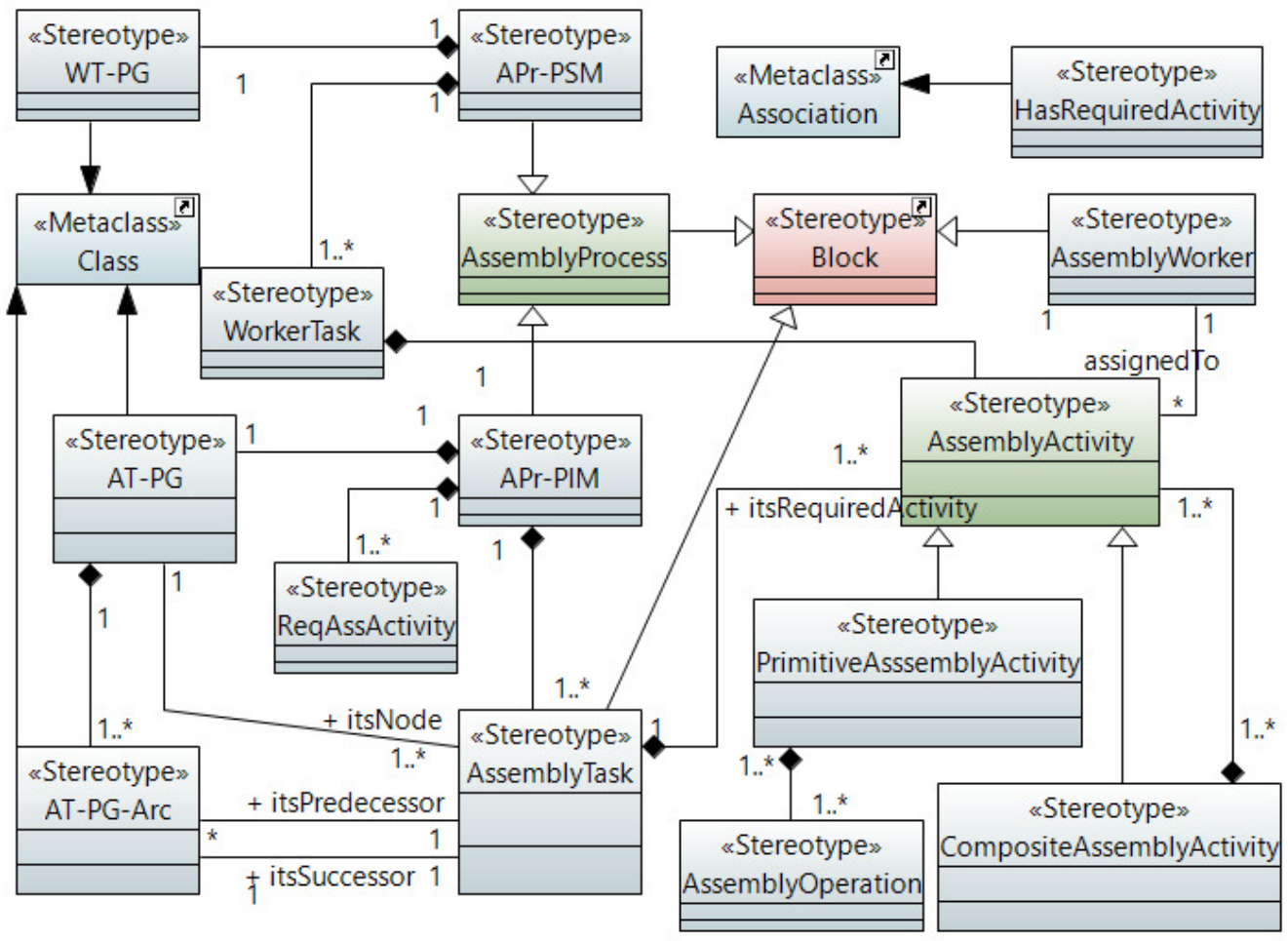

Fig. 8. The Assembly Process meta-model as a UML profile (core part).

The AT-PG of the APr-PIM defines the solution space of all the acceptable assembly scenarios that may be adopted in order for the corresponding Assembly to be realized. It is composed of nodes that represent the assembly tasks of the APr-PIM and arcs that represent precedence relations among ATs. Every AT-PG has a *MasterInitialTask* (MIT) and a *MasterFinalTask* (FT). Optionally, it has a *BranchInitialTask* (BIT), and *BranchFinalTask* (BFT) for each *BranchSubAssembly* captured on the PSM. The MIT is composed of activities required to start the assembly processes, for example the activity to hold the base part (*itsBasePart*) of the *MasterSubAssembly*. The FT is composed of activities required to finalize the product and remove it from the assembly or sub-assembly line. The BIT and the BFT, which is optional, are defined in an analogous way to the one of MIT and FT definitions for each *BranchSubAssembly*. As for the Gregor chair case study, the FT is composed of the activities Release and PickAndPlace of the product to the conveyor belt.

The service-oriented job assignment adopted in this framework is based on provided and required services. A provided service refers to an assembly activity that the assembly worker (human or machine) may perform in response to a request to offer the specific service to its environment. The term quality of service characteristics (QoSs) is used to refer to the quality characteristics of the assembly activity. The sequence of moves and the basic processes as defined at MTM-SD, e.g., grasp and release, and put in place and MTM-UAS/MEK, e.g., grasp and put in place (Almeida and Ferreira 2009), are considered as examples of provided services. However, we must note that more descriptive names should be given to the assembly activities not just a composition of the used assembly operations. Any assembly activity that is a constituent part of an assembly task that constitutes an APr-PIM, is considered as a required service. Any primitive, no further decomposed operation, that may be performed by an assembly worker during the assembly process, is considered as an assembly operation. Characteristic examples of Assembly operations include the 5 basic movements defined in MTM, i.e., Reach, Grasp, Move, Position, and Release (Almeida and Ferreira 2009). A worker may expose services to its environment that constitute the access points for triggering the execution of the corresponding assembly activities. Assembly activities are either composite or primitive. An assembly operation can be exposed as a service only in the form of a primitive assembly activity (*PrimitiveAssemblyActivity*).

The part of the behavior of an AT that is assigned to a specific worker is defined as Worker Task (WT). Thus, an AT can be allocated to one worker (in this case the WT number is the number of the AT) or decomposed into parts with each one assigned to different worker (in this case the WTs are numbered with the number of the AT and their order in the decomposition, i.e. WT1.1 is the first WT of AT1. The WT is specified in terms of the assembly activities of the AT. The assembly activity is the unit of distribution of the assembly work to assembly workers. Precedence relations among the WTs are expressed in the WT-PG. The APr-PSM and its sub-processes are modelled using the UML activity diagram. The concept of swimlane is used to capture the assignment of assembly activities to assembly workers. A precondition in the activity diagram defines the location in which the worker should be to execute the specific worker task. The worker has to move to the right location before executing the WT. The execution of the WT is triggered when the part or subassembly, that the WT operates on, is in a proper location and possibly a proper status.

### *5.2 Plant independent modelling of assembly processes and tasks*

Assembly processes utilize directly or indirectly functionality provided by p-CPuSs, as well as computational microservices, to provide a higher layer functionality required at the process level of the plant, as shown in Figure 3. Thus, assembly and sub-assembly processes are modelled as c-CPuS, i.e., as compositions of worker tasks, adopting the orchestration and/or the choreography pattern. Both patterns have been implemented and examined in the prototype implementation of the Gregor chair case study. Choreography matches the semantics of the decentralized networked control systems. In this case, there is no centralized control that captures the coordination logic of the WTs during the assembly or sub-assembly process. A state-of-the-art review on this subject is given by Bakule (2014). Worker tasks are also modelled as c-CPuSs, i.e., as compositions of assembly activities following the orchestration pattern. Chunks of assembly functionality at the plant process layer, involving more than one CPuS are also modelled as CPuSs to have a modular and flexible assembly process layer implementation. For example, the CPuS that implements the assembly work required for the assembly task 1 (AT1) of the Gregor chair case study is a classic example of a composite CPuS.

Based on the above scenario, the assembly engineer defines the APr-PIM, i.e., they specify the assembly process in a plant independent manner. PIM specifies the assembly activities that should be performed without using specific workers or any info related to the plant configuration. For example, operations such as move, and transfer, have to do with the assembly platform configuration and are not included in the PIM model. These operations will be inserted in the model in the next phase when the PIM will be transformed to a plant-specific model (PSM), i.e., during the time a requested assembly activity spec of the PIM is resolved to a specific assembly activity provided by a specific worker.

### *5.3 The construction of plant independent model for the assembly process*

The Assembly Engineer constructs the APr-PIM for a specific product based mainly on the Product's Structural Model. This is a three-step process:

1. Identification of ATs.
2. Construction of the AT-PG.
3. Specification of the required assembly activities.

#### *A) Identification of ATs*

The set of ATs of an APr-PIM is derived from the corresponding PSM based on the following rules.

*Rule 1*: One AT per liaison
An assembly task is defined for each «*curDclLiaison*» or «*LoDclDefinedL*» of the APr-PIM except for the cases where rule 2 is applied. An AT is not defined for a «*HiDclrealisedL*» liaison.

*Rule 2*: One AT for more than one liaisons
More than one liaisons should be assigned to the same AT in the case one of the following conditions applies:

a) A part has more than one liaisons with other parts which at the time of realization of the liaison happen to be parts of the sub-assembly on which the part is going to be assembled. In this case, all these liaisons are associated to the same assembly task.
b) A *BranchSubAssembly* has more than one liaisons that connect its parts with parts of the *MasterSubAssembly* to which it is going to be assembled. In this case, all these liaisons are associated to the same assembly task.

Figure 9 presents the PSM of the Gregor Chair as it has been constructed using the Papyrus UML tool and the PSMM profile. In Figure 10, information regarding the master and one branch sub-assembly, i.e., the one corresponding to the *UpperSubAssembly* composite part has been depicted. ATs of the APr-PIM are also shown on this figure.

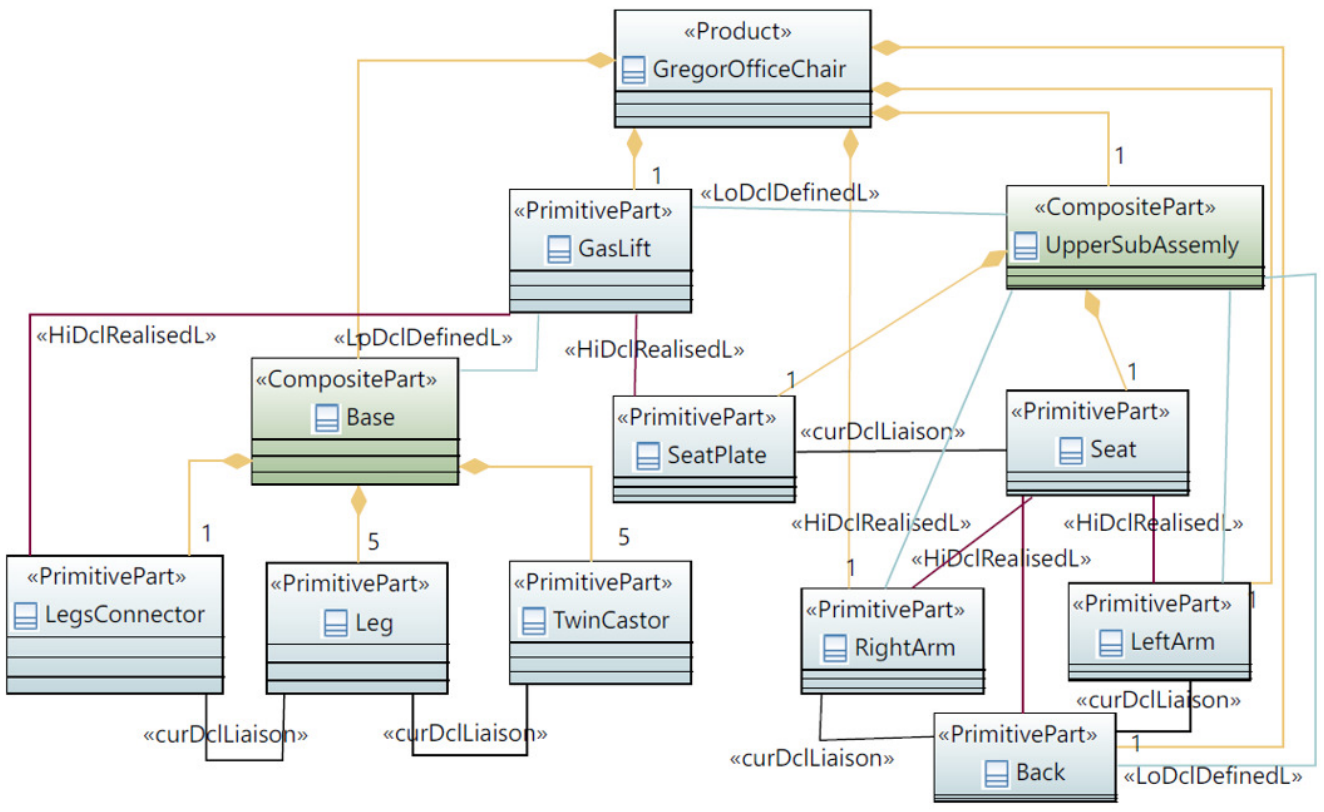

Fig. 9. The structural model (PSM) of the Gregor chair.

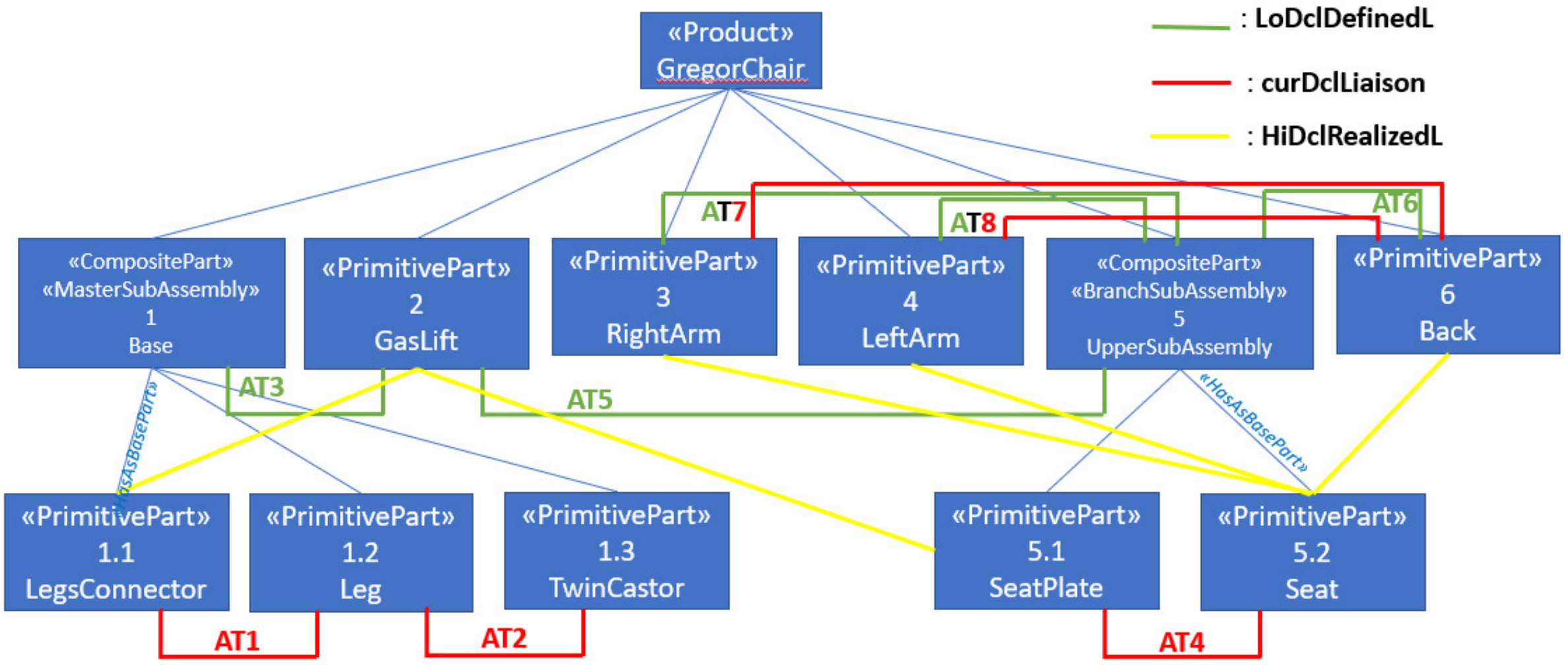

Fig. 10. The refinement of the PSM of the Gregor chair with information to facilitate the generation of the APr-PIM.

*B) Construction of the AT-PG.*

The construction of the AT-PG is based on the following rules:

*Rule 1*: MIT construction rule

The BasePart (itsBasePart) of the MasterSubAssembly of the PSM results in the construction of the MIT of the AT-PG.

*Rule 2*: BIT construction rule

The BasePart (*itsBasePart*) of each *BranchSubAssembly* of the PSM results in the construction of BIT for the AT-PG.

*Note*: The same rule is applied during the refinement process of the APr-PIM to get an APr-PSM if the assembly engineer decides to assemble a composite part of the PSM independently of the *MasterSubAssembly*. In this case, the base part (the endpoint of the *HasBasePart* association) of the composite part should be identified on the PSM.

*Rule 3*. The arcs generation rule

This rule is given in the form of an algorithm. For the generation of the arcs of the AT-PG, the definePG-Arcs algorithm is executed with the Product instance of the APr-PIM as *curNode*.

```
For each curNode.child
    if curNode.child.type is CompositeChildComponentPart
        definePG-Arcs(curNode.child)
    if curNode.child.type is PrimitiveChildComponentPart
        processChildrenLiaisons(curNode.child)
processChildrenLiaisons(curNode)
```

where

*processChildrenLiaisons* is defined as follows

*processChildrenLiaisons* of *curNode*

define the arcs among the ATs that correspond to liaisons among children of the *curNode* based on the order of the liaisons (see order property of Liaison stereotype and the semantics of the *MasterSubAssemblystereotype*), and *CompositeChildComponentPart* and *PrimitiveChildComponentPart* are defined in an analogous way with CCAP and PCAP used in the assembly process definition expression.

Fig. 11a presents the AT-PG of the Gregor chair case study and Figure 11b presents the refinement of the AT-PG to capture decisions regarding branch sub-assemblies and the pruning of the solution space. Figure 12 presents the specification of the assembly task AT1 of the case study.

## 6. THE CPuS-IoT FRAMEWORK

Several notations are used for service orchestration with the goal to be usually twofold, flexibility and responsiveness. The objective of the CPuS-IoT framework is to fulfill both requirements. Responsiveness is addressed at the p-CPuS level by encapsulating the mechanical unit control and coordination logic in the microservice level, i.e., in the p-CPuS, close to the physical plant unit. Flexibility is achieved by several means. As a first step, assembly processes are implemented as dynamically deployable c-CPuSs, which are executed in a microservice container that supports run-time reconfiguration, e.g., OSGi or node.js, both experimented in our prototype implementation. Moreover, assembly processes may be defined without any reference to specific services provided by the assembly platform. This allows an assembly process, i.e., a c-CPuS, to dynamically acquire at deployment and even at run-time, the available assembly workers or other artefacts, which are required to fulfill its goals, i.e., to execute the requested assembly activities. The adopted approach establishes the basic requirements that characterize the system as evolvable. Assembly and subassembly processes as well as worker tasks are generated on demand based on the product variant model and deployed automatically on the assembly platform, exploiting the corresponding features of containers, for the assembly of the corresponding product variant. This characterizes the CPuS-IoT approach as goal-driven.

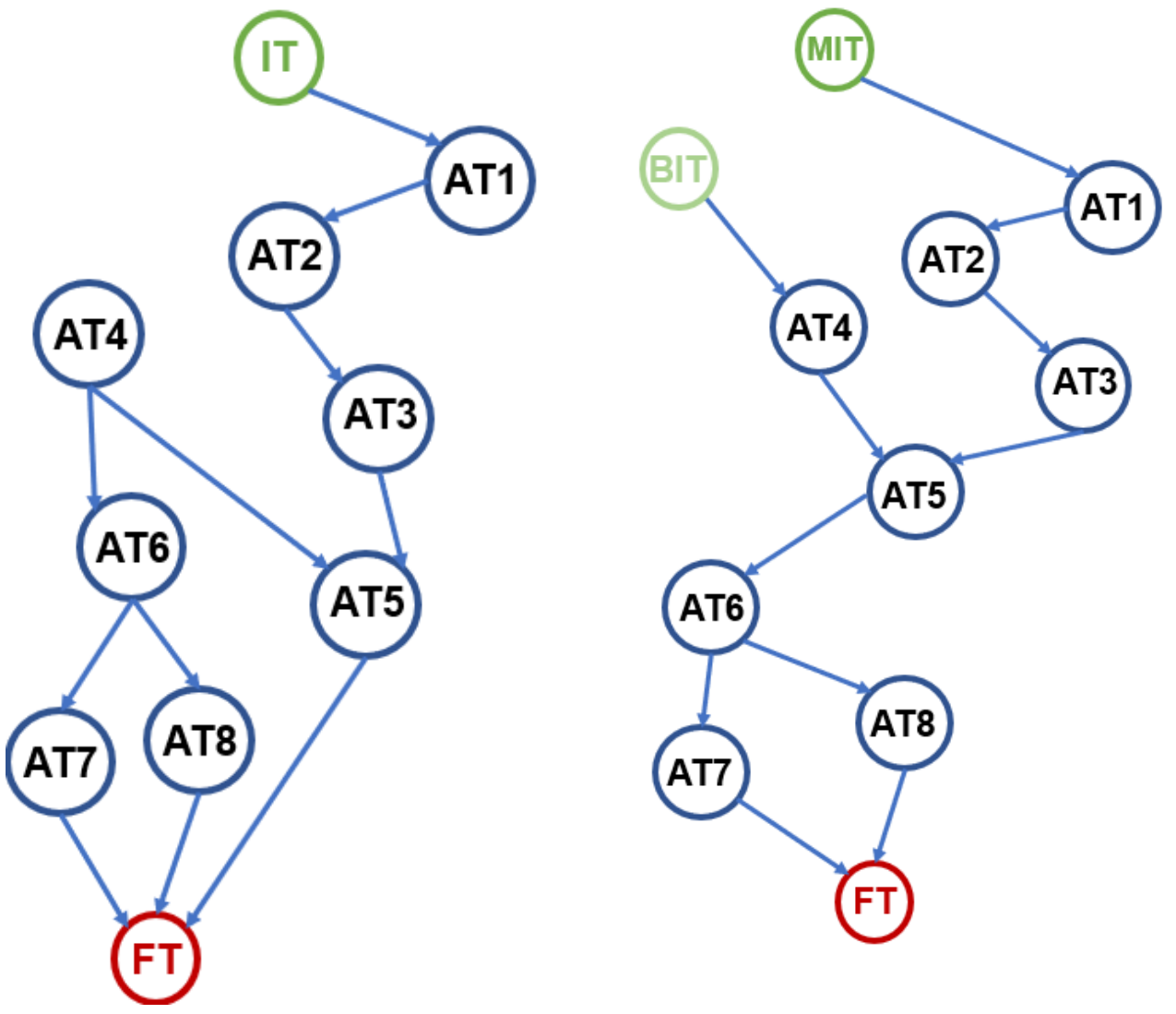

(a) (b)

Fig. 11. Assembly Task precedence graphs for the Gregor Chair case study. (a) Initial AT-PG extracted from the Gregor chair structural model. (b) Refinement of the initial AT-PG to capture decisions on branch sub-assemblies.

```
For each Leg
  PickAndInsert Leg into LegsConnector
  Rotate Fixture

For each Leg
  ScrewPickAndFasten on the ThreadedHoleEndPoint
  Rotate Fixture
```

Fig. 12. Description of the Assembly Task AT1 that corresponds to liaison Lp1.1p1.2.

### *6.1. The PIM to PSM Transformation process*

The transformation of the APr-PIM to the APr-PSM can be performed manually by the control engineer or automatically by the framework. The framework supports this operation through a service discovery mechanism, as shown in Figure 13 which captures the framework infrastructure that is related to the transformation of APr-PIM to APr-PSM. This mechanism can be utilized either for a static assignment of provided services or a dynamic one. In the case of dynamic assignment of services, the system will check for the availability of primitive CPuSs providing the physical operations and satisfying the requested service specs and the prerequisites of using them. Then, it will instantiate the process c-CPuS reserving the required CPuSs. An alternative is for the system to postpone the reservation of resources up to the time they are required. This functionality of the framework supports a better use of the platform's resources and allows a more flexible process implementation. The c-CPuS description is a prerequisite for the realization of the APr-PIM to APr-PSM transformation.

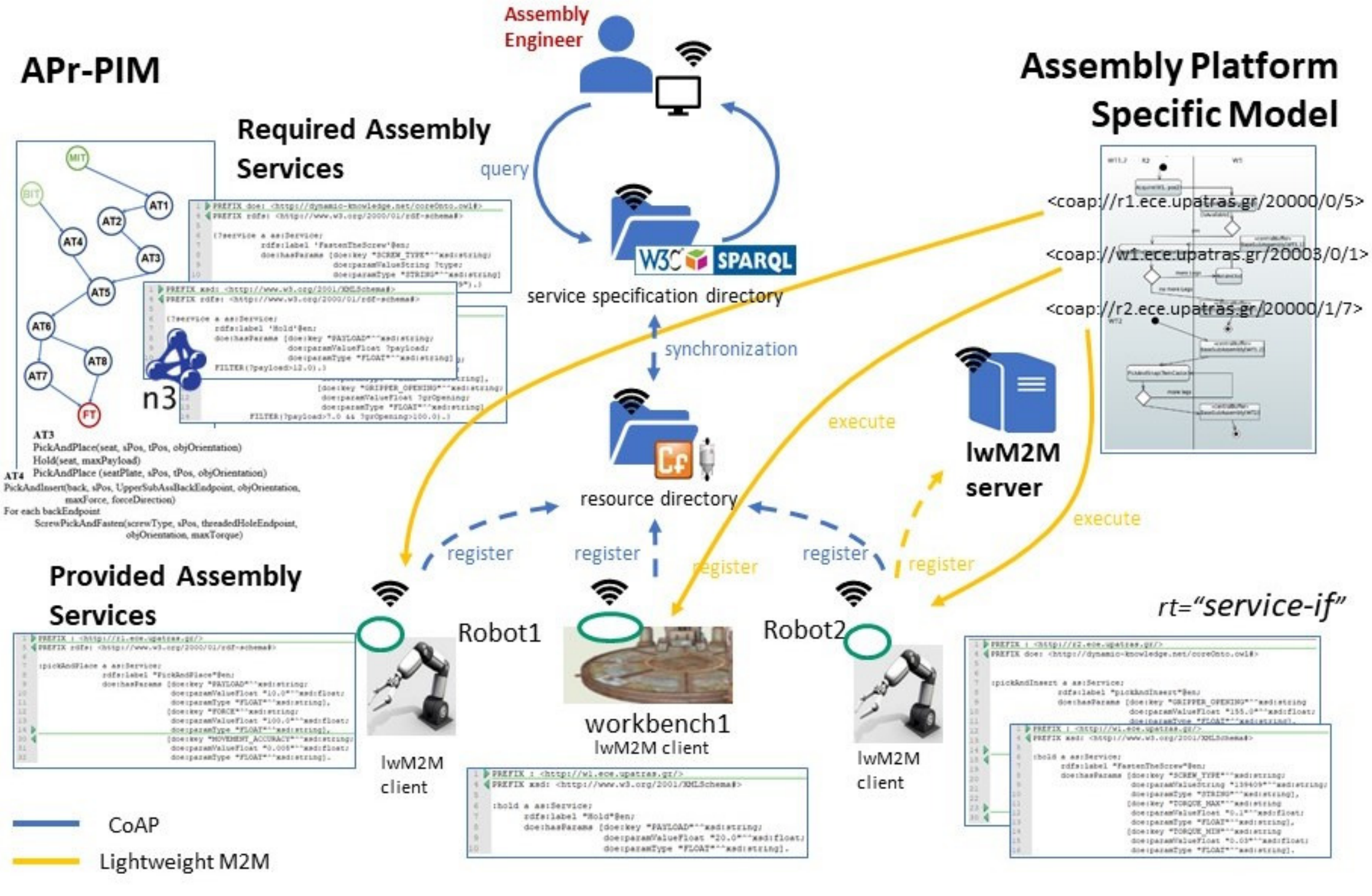

Fig. 13. A goal-driven service composition approach for the Assembly Process Model: From APr-PIM to APr-PSM.

### *6.2. Description and discovery of CPuS*

An assembly worker, such as the robot R1, exposes its provided services, e.g., *PickAndInsert* and *ScrewPickAndPlace*, as resources. These services will be used for the realization of an AT's liaisons, as for example the tapering and screw fitting needed in AT1 of the GC case study. For the framework to support service discovery during development time but also during run-time, an efficient description is required for the provided services. For the description of the provided services of the p-CPuS the Core Ontology (https://wiki.tut.fi/DOE/CoreOntology), introduced by *Lanz et al.* 2018, is used. The IPSO smart object description has been extended with the description of the provided services as well as the services' states expressed in Notation 3 or RESTdesc.

Notation 3 (N3) is an assertion and logic language that extends the RDF by adding formulae, variables, logical implication and functional predicates (https://www.w3.org/TeamSubmission/n3/). It is based on Statements, which are triples consisting of a Resource, a Property and the value of the Property, represented by URIs and serving as subject, predicate and object, respectively. For example, the triple *local:pickAndPlace a as:Service* defines *pickAndPlace* as a service (*a* serves as an N3 abbreviation for the *rdf:type* property) and the *rdfs:label* instance of Property is used to define a human-readable name for the resource. Properties are also used to express attributes of a resource or a relationship between two resources.

RESTdesc is a machine-interpretable functional service description format for REST APIs (Verborgh *et al.,* 2012) that exploits HTTP vocabulary and N3 to enable the machine to discover and consume Web services based on links (Verborgh *et al.,* 2011). RESTdesc descriptions include a set of preconditions and a set of postconditions, indicating that if the preconditions in the antecedent are true for a specific substitution of the variables, then an HTTP request will be feasible for the realization of a service by using URIs or request bodies associated with the same substitution. A mechanism that allows RESTdesc to capture states was introduced by Mayer *et al.* (2014) and extended by Kovatsch *et al.* (2015), enabling the description of service states. N3 statements may provide information about the functionality of a service and information about Quality of Service (QoS) characteristics. For example, all holding services provided by different workbenches should have a common label "Hold", but possibly different levels of QoS regarding the maximum allowed payload that can be hold. Figure 14 captures part of the description for a pick-and-place provided service which is labelled accordingly and has specific QoS characteristics e.g., it accepts only input objects that require a gripper opening of 155mm at most, weight up to 10kg and are placed within a range of 1300 mm.

```
PREFIX : <http://r1.ece.upatras.gr/>
PREFIX rdfs: <http://www.w3.org/2000/01/rdf-schema#>

:pickAndPlace a as:Service;
              rdfs:label "PickAndPlace"@en;
              doe:hasParams [doe:key "PAYLOAD"^^xsd:string;
                             doe:paramValueFloat "10.0"^^xsd:float;
                             doe:paramType "FLOAT"^^xsd:string],
                            [doe:key "FORCE"^^xsd:string;
                             doe:paramValueFloat "100.0"^^xsd:float;
                             doe:paramType "FLOAT"^^xsd:string],
                            [doe:key "FORCE_ACCURACY"^^xsd:string;
                             doe:paramValueFloat "5.5"^^xsd:float;
                             doe:paramType "FLOAT"^^xsd:string],
                            [doe:key "GRIPPER_OPENING"^^xsd:string;
                             doe:paramValueFloat "155.0"^^xsd:float;
                             doe:paramType "FLOAT"^^xsd:string],
                            [doe:key "RANGE"^^xsd:string;
                             doe:paramValueFloat "1300.0"^^xsd:float;
                             doe:paramType "FLOAT"^^xsd:string].
```

Fig. 14. N3 description of *PickAndPlace* CPuS of assembly worker R2.

### *6.3. The prototype implementation of the CPuS-IoT framework*

The CPuS-IoT framework supports the discovery of assembly services using a service repository where the provided assembly activities of the assembly platform are automatically registered by their hosting workers. The CoRE resource directory (Shelby *at al.*, 2018) defined by the IETF CoRE Working Group is adopted in this work. It enables methods for discovering a resource directory (RD), as well as registering and looking up resource descriptions. It targets resource-constrained devices used in M2M applications and surpasses the problems that direct discovery imposes, by employing an RD which hosts accessible descriptions of resources held on servers The californium.tools repository (Shelby *at al.*, 2018) is used as a Cf-RD resource directory implementation to be aware of the devices and services of the assembly platform.

Each device hosting services for assembly activities accesses the RD and sends a POST request through the registration interface. The message payload contains the list of resources offered by the device in the CoRE Link Format as well as the semantic and dynamic state descriptions of the provided resources. The RD lookup and update mechanisms allow the search and discovery of the exposed resources and the access to up-to-date information concerning resource descriptions. In the Gregor chair case study, the p-CPuSs register to the RD once activated and publish lists of provided services, e.g., *pickAndPlace*, *screwPickAndFasten* and *hold*, along with their N3 or RESTdesc descriptions. The development environment or an agent, for the case of operation-time discovery, accesses the descriptions and looks for resources that offer the desired functionality for the realization of an assembly task, such as the realization of the screw fit joint between the seat plate and seat primitive parts, i.e. AT4. The SPARQL query language for RDF enables the filtering of services which meet the process requirements. For example, during the assembly task AT4, the control engineer performs queries to identify *pickAndPlace* services with specific QoS characteristics, to specify and potentially utilize the entities that provide these services. Figure 15 shows a SPARQL query for discovering assembly services that pick and position payload with maximum allowed weight greater than 7kg by using a finger gripper that spreads up to 100mm.

```
PREFIX xsd: <http://www.w3.org/2001/XMLSchema#>
PREFIX rdfs: <http://www.w3.org/2000/01/rdf-schema#>

{?service a as:Service;
          rdfs:label 'PickAndPlace'@en;
          doe:hasParams [doe:key "PAYLOAD"^^xsd:string;
                         doe:paramValueFloat ?payload;
                         doe:paramType "FLOAT"^^xsd:string],
                        [doe:key "GRIPPER_OPENING"^^xsd:string;
                         doe:paramValueFloat ?grOpening;
                         doe:paramType "FLOAT"^^xsd:string]
          FILTER(?payload>7.0 && ?grOpening>100.0).}
```

Fig. 15. Example query for the discovery of PickAndPlace assembly service with specific QoS.

Figure 16 provides an indication of the communication and processing overhead introduced by the proposed framework for triggering the execution of a service of an assembly worker. More specifically, it captures the round-trip time for the EXECUTE operation of the LwM2M protocol that is utilized for triggering the execution of a CPuS in our prototype implementation.

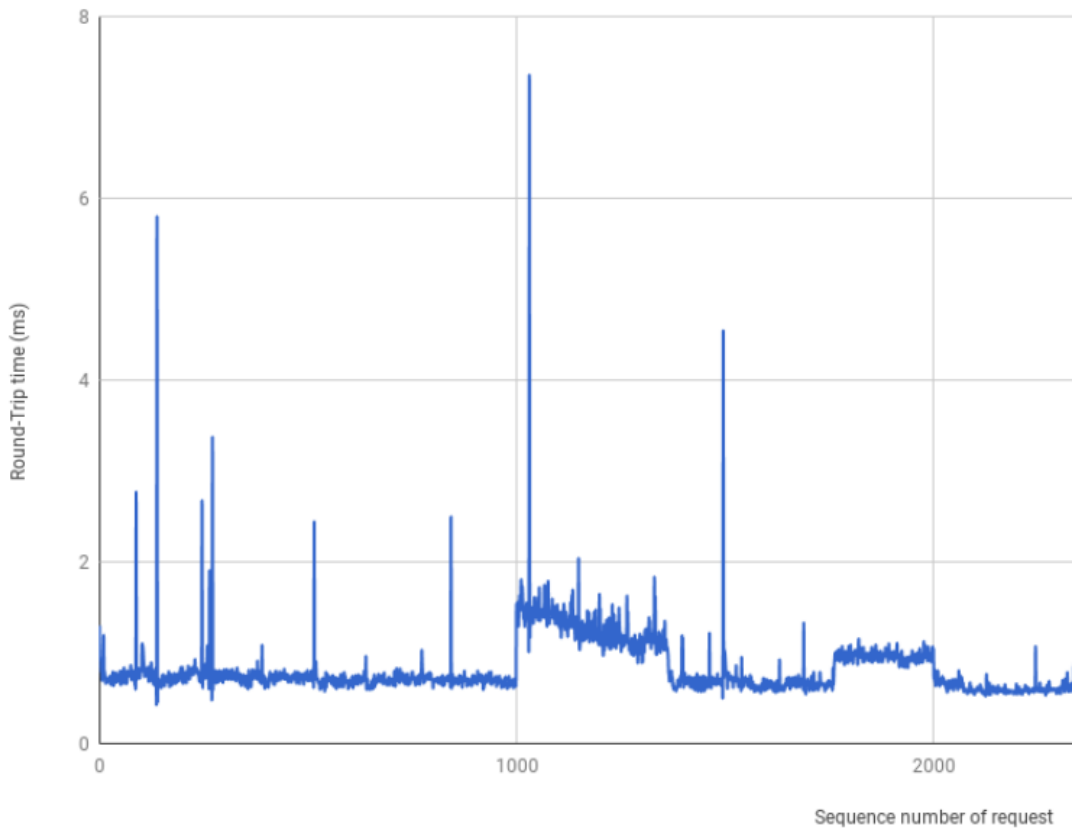

Fig. 16. The round-trip time for the EXECUTE operation of the LwM2M IoT application layer protocol.

## 7. CONCLUSIONS

The requirements for mass customization increase the complexity of manufacturing assembly systems. Legacy assembly systems designed with the objective of mass production, should be replaced by evolvable ones exploiting current advances in IT. This transformation is not an easy task. Specific approaches and frameworks are required to effectively integrate state-of-the-art technologies to address the challenges in this domain. Towards this direction, we have presented in this paper a) the key concepts of a cyber-physical microservice and IoT-based approach and framework for evolvable assembly systems of the 4th

Industrial revolution, and b) an approach for the product's structural modelling process and its use for the automatic construction and run-time evolution of the assembly process.

Assembly workers as well as other artefacts involved in the assembly process are transformed to smart entities (cyber-physical entities), which are represented in the assembly system platform level as IoT-compliant entities exposing their properties and functionalities as cyber-physical microservices (CPuSs). The number of CPuSs offered by a structural component of the assembly platform is dependent upon its complexity. Simple components offer just one CPuS, while complex ones may offer more than one CPuS. A bottom-up approach has been presented for the assembly engineer to design the assembly system following an MDE approach that exploits both the orchestration and choreography pattern in service composition. By adopting web-based representations of models and meta-models, that capture the domain knowledge, as well as appropriate inference engines, significant parts of the design process of the assembly system can be semi or even fully automated. Furthermore, this representation is the infrastructure for the dynamic, without human intervention, reconfiguration of the assembly process to the requirements of the specific product variant. Based on this, the presented approach can be characterized as belonging to the goal-driven service composition paradigm. We claim that this framework provides the basics for a common vocabulary to be defined as well as the infrastructure that is required for the implementation of various assembly algorithms. Even though the paper focuses on the assembly systems domain, most of the key concepts apply to the manufacturing domain in general.

We are currently working on a) a more detailed modelling of the assembly platform, b) on the semi-automation of various parts of the design process of the assembly system exploiting semantic web for assembly service discovery and composition, and c) the use of real-time containers as artefacts to enable CPuSs to address real-time constraints inherent in many manufacturing structural components. Work in progress involves also the demonstration, on the test bed, of the evolvability features of the CPuS-IoT framework that includes the demonstration of the goal-driven nature of the CPuS framework concerning service composition. Future work will focus on a detailed definition of the semantics of the CPuS that will allow the formal verification of an assembly process defined based on the orchestration and/or choreography patterns of service composition. Further development will address the use of RESTdesc that additionally to the RDF provides the hypermedia links needed to access the resources, enhancing decoupling between p-CPuSs offered by assembly workers and plant processes.

**ACKNOWLEDGMENTS**

The Authors would like to thank the anonymous reviewers for their comments that resulted in an improved version of the paper.